\documentclass[twocolumn,aps,showpacs]{revtex4}
\usepackage{graphicx}
\usepackage{bm}

\begin{document}

%
%

\title{Efficiency of electrical manipulation on two-dimensional topological insulators}

\author{M. Pang and X. G. Wu}

\affiliation{SKLSM, Institute of Semiconductors, Chinese Academy
of Sciences, Beijing 100083, China}

\begin{abstract}

We investigate the efficiency of electrical manipulation on two-dimensional topological insulators
by considering a lateral potential superlattice on the system.  The electronic states under various
conditions are examined carefully.  It is found that the dispersion of the mini-band and the
electron distribution in the potential well region display an oscillatory behavior as the potential
strength of the lateral superlattice increases.  The probability of finding an electron in the
potential well region can be larger or smaller than the average as the potential strength varies.
This indicates that the electric manipulation efficiency on two-dimensional topological insulators
is not as high as expected, which should be carefully considered in designing a device application
that bases on two-dimensional topological insulators. These features can be attributed to the
coupled multiple-band nature of the topological insulator model. In addition, it is also found that
these behaviors are not sensitive to the gap parameter of the two-dimensional topological insulator
model.

\end{abstract}


\maketitle

%
%

\section{Introduction}

Topological insulator has received widespread attentions in recent years because of its exotic
electronic properties which suggest wonderful prospects both in the fundamental study and
industrial applications \cite{Physics.1.6, qi:33, RevModPhys.82.3045, Moore10Nature}.  Its novel
feature lies on the observation that when it is of a finite size it has a linearly dispersive
surface states (edge states in the two-dimensional system case) in the bulk gap
\cite{Konig02112007, Hsieh08Nature, Analytis10NatPhys}.  The robustness of the surface states
against disorder scattering makes topological insulator a promising new member of materials useful
in the future information technology. In practical applications, electric manipulation of the
carriers by voltage gates provides a basic important mechanism for the functioning of the devices.
It thus becomes a crucial tasks is to learn how to control the behavior of electrons in topological
insulators using electric fields. Various research concerning this issue have been done in the
literature \cite{NJP083058, PhysRevB.81.235323, PhysRevB.81.115322, PhysRevB.83.081402,
PhysRevB.83.165304, PhysRevB.84.161301, PhysRevB.85.115434, PhysRevB.85.125309, PhysRevB.85.235131,
PhysRevB.86.165418, PhysRevLett.104.246806, PhysRevLett.106.057205}. If the edges where the edges
states localize can be created by electric voltage gates, namely, carriers in the system can be
confined spatially by electric potential, it would greatly enhance the utilization of topological
insulators as a building block of future devices. Actually, such design has been proposed
\cite{PhysRevB.83.081402} where the possibility of edges created by voltage gate is implicitly
assumed without a serious and careful scientific justification.

Theoretically, the effective model put forward in Reference \cite{Bernevig15122006} has been widely
used to describe the two-dimensional (2D) topological insulator \cite{NJP083058,
PhysRevLett.101.246807, PhysRevLett.102.136806, PhysRevLett.106.206802, PhysRevLett.107.086803,
PhysRevB.79.241306, PhysRevB.81.241303, PhysRevB.81.115322, PhysRevB.81.235323, PhysRevB.83.081402,
PhysRevB.85.115433, PhysRevB.85.125309, PhysRevB.85.155308}. In this model, one deals with a
coupled two-component system (more generally, when the spin-orbit interaction is considered, it is
a four-component system). The model resembles the Dirac equation describing a Dirac fermion which
is known to exhibit the Klein tunneling phenomenon \cite{Dombey199941} and consequently a potential
barrier may be unable to confine the fermion motion spatially.

Therefore, it will be interesting and meaningful to examine the influence of a spatially defined
potential on the 2D topological insulator, especially the confinement of electron wave functions
induced by the externally applied potential.  The present paper addresses this very important
problem. We simplify the issue by considering a lateral potential superlattice structure made of a
2D topological insulator with spatially periodic, square shaped, electric potential barriers
created by electric gate voltage. The single-electron energy dispersion and wave function will be
calculated.  The spatial dependence of wave function will be investigated when the lateral
superlattice potential is varied.

The paper is organized as follows: In section \ref{sec.theo}, the theoretical formulation is
presented. Section \ref{sec.dis} contains our calculated results and their discussions. Finally, in
the last section, a summary is provided.

\section{Formulation and calculation}\label{sec.theo}

The 2D topological insulator is modelled by the following well-known Hamiltonian
\cite{Bernevig15122006}
\begin{equation}\label{effmodel}
   \left(
   \begin{array}{*{20}c}
   {h( {\bf k} )}  &  {0}  \\
   {0}  &  { h^*(-{\bf k}) }  \\
   \end{array}
   \right),
\end{equation}
where ${\bf k}=(k_x,k_y)$. $h({\bf k})=\epsilon({\bf k})+\sum_\alpha d_\alpha({\bf
k})\sigma_\alpha$ with $\sigma_\alpha$ the Pauli matrix.  $\epsilon({\bf k})=C-D(k_x^2+k_y^2)$,
$d_1=Ak_x$, $d_2=Ak_y$, and $d_3=M-B(k_x^2+k_y^2)$.  $A$, $B$, $C$, $D$, and $M$ are parameters
determined by the structure of the quantum well \cite{Bernevig15122006}.  The parameter $C$ gives
the zero point of energy and in this paper we can safely set it to zero for simplicity. When there
is no coupling between spins (not the true spin, just for labeling convenience
\cite{Bernevig15122006}), the lower $2\times 2$ block $h^*(-{\bf k})$ and the upper block $h({\bf
k})$ are decoupled.  They have the same energy spectrum and the corresponding wave functions are
connected via an unitary transformation of time reversal operator. Thus in this paper we will limit
ourselves to the analysis of upper block $h({\bf k})$ only.

The topological insulator is in the $xy$ plane, and the lateral potential superlattice is assumed
to be along the $x$ axis and uniform in the $y$ direction.  The Hamiltonian describing the lateral
superlattice is assumed to be given by $H=h({\bf k})+V(x)$. The potential $V(x)$ is diagonal and
periodic in $x$.  For simplicity, the square barrier potential is assumed.  In one unit cell $x \in
[0, L]$, $V(x)=0$ for $x\in [R/2, L-R/2]$, and $V(x)=V_0$ otherwise.  $L$ is the period of the
lateral superlattice and $R$ is the width of the potential barrier in one unit cell.  The lateral
superlattice could be implemented by periodically depositing metal strips on the surface of the 2D
topological insulator with the help of modern lithography technology.  The barrier potential height
can be varied by adjusting the bias voltage applied to the metal strips.  The 2D topological
insulator model is derived from a multiple band envelope function model \cite{Bernevig15122006}.
Thus the assumption of a diagonal $V(x)$ is reasonable and consistent with the model, actually,
similar kind of potential profile has been used in graphene \cite{JPCM-465302}.

The system is translational invariant in the $y$ direction, and $k_y=p_y/\hbar$ is a good quantum
number.  But because of the presence of potential $V(x)$, there is no translational invariance in
the $x$ direction, and we follow the Peierls substitution to replace $k_x$ in Equation
(\ref{effmodel}) by the differential operator $-{\rm i}\partial_x$. The explicit form of the
Hamiltonian is thus as follows.
\begin{equation}\label{effmodel2}
H = \left(
   \begin{array}{*{20}c}
   {M + d_+k^2 + V(x)}  &  {Ak_-}  \\
   {Ak_+}  &  {-M + d_-k^2 + V(x)}  \\
   \end{array}
   \right),
\end{equation}
with $d_+=-(B+D)$, $d_-=B-D$, $k^2=k_+k_-$, $k_+=-i\partial_x+ik_y$, and $k_-=-i\partial_x-ik_y$.

The wave function of the above Hamiltonian can be written as
\begin{equation}
   \psi(x,y) =
   {{{\rm e}^{{\rm i}k_y y}}\over{\sqrt{L_y}}}
   {\rm e}^{{\rm i}q_x x}
   \left( \matrix{
   u^{\rm u}(x) \cr u^{\rm l}(x) \cr }
   \right),
\end{equation}
with $L_y$ the system size in the $y$ direction and $q_x$ the mini-band wave vector. $u^{\rm
u}(x)=u^{\rm u}(x+L)$ and $u^{\rm l}(x)=u^{\rm l}(x+L)$ give the upper and lower components of the
periodic part of the wave function.  The wave vector $q_x$ should be limited to the first Brillouin
zone $q_x \in [-\pi/L, \pi/L]$.  In the present paper, the plane wave expansion approach is used
and sufficient number of plane waves are used to achieve the necessary accuracy for the energy
levels.  The wave function $u^{\rm s}(x)$ can be written as $u^{\rm s}(x) = L^{-1/2} \sum_n c^{\rm
s}_n {\rm e}^{{\rm i}2\pi nx/L}$ where $\rm s$ stands for $\rm u$ and $\rm l$, respectively.   The
wave function $u^{\rm s}(x)$ is normalized to unity in one unit cell.

The model for the 2D topological insulator contains many parameters.  In the present paper, we will
mainly consider the influence of the gap parameter $M$, as its sign has a profound effect on the
electronic property when the 2D topological insulator has a finite size
\cite{PhysRevLett.101.246807, Bernevig15122006}. The numerical values of other parameters $A$, $B$,
and $D$ are taken from reference \cite{Bernevig15122006}.  The influence of the strength of
superlattice potential $V_0$ will also be studied.

\section{Results and discussions}\label{sec.dis}

In the absence of periodic potential, the bulk 2D topological insulator approximately has two bands
\cite{Bernevig15122006}.  The introduction of the lateral superlattice potential $V(x)$ leads to a
multiple energy band structure. The eigen energy of the Hamiltonian $H$ becomes $E_i(q_x, k_y)$
with $i$ the band index.  Extensive numerical works are carried out and qualitatively the same
results are obtained.  In the following we will show calculated results for a typical case of
$L=20$ nm and $R=L\times 0.5123456$. The ratio $R/L$ is chosen to avoid possible length
commensurate artifact and other ratios are also used.

It is found that, as $V_0$ increases (decreases), each band energy $E_i(q_x, k_y)$ increases (decreases) as a
whole, roughly with an amount $V_0R/L$, the averaged potential energy in one unit cell.  Furthermore, for a
given $V_0$, if one takes the averaged potential $V_0R/L$ as the energy zero point, then the $m$-th band above
or below this averaged potential energy will approximately has its largest expansion coefficient $|c^{\rm s}_n|$
at $|n|\sim |m|$ ($n=0,\pm 1, \pm 2, ...$). $|c^{\rm s}_n|^2$ is shown in Figure 1, for a typical case of
$q_x=0.01$ nm$^{-1}$ and $k_y=0$ with $M=-6.86$ meV.  An almost linear dependence can be obviously seen.
This shows that the bands near the averaged potential are more ``ground state'' like as it is less oscillatory
compared to the bands more far away from the averaged potential.  It is interesting to point out that when the
sign of $M$ changes, the behavior shown in Figure 1 is qualitatively the same, despite the fact that when $M>0$
there is no edge state when the system is of a finite size.
\begin{figure}[ht]
\includegraphics[width=6cm,height=9cm,angle=-90]{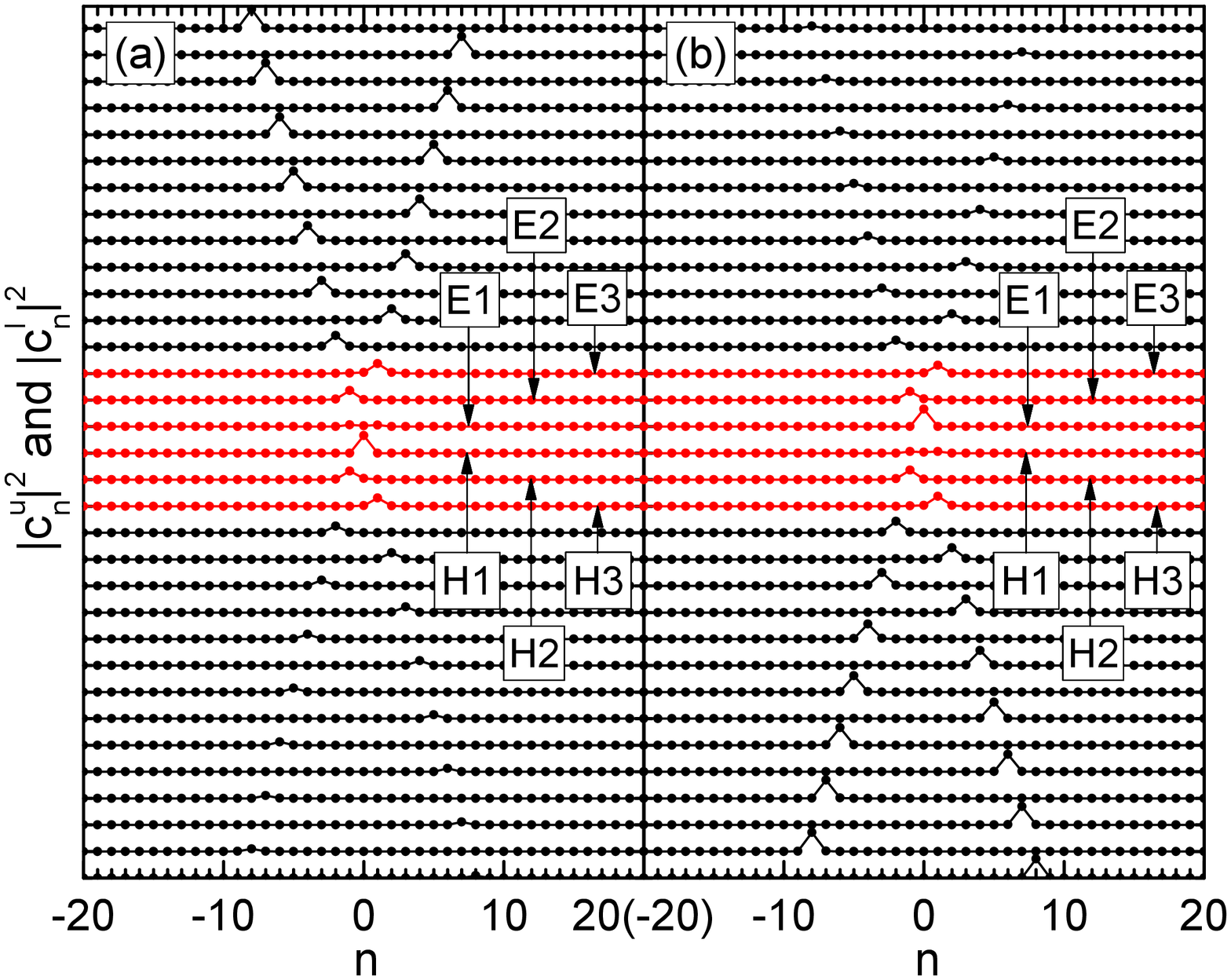}
\caption{(Color online)
Square of the Fourier expansion coefficients $|c_n^{\rm u}|^2$ (panel (a)) and
$|c_n^{\rm l}|^2$ (panel (b)) of wave functions for energy
levels near the averaged potential $V_0R/L$.  $L=20$ nm, $q_x=0.01$ nm$^{-1}$,
$k_y=0$, $V_0=0.1$ eV, and $M=-6.86$ meV.  The coefficients are shown as dots and
lines connecting the dots are guide to the eye.
For clarity, the coefficients for different energy levels are shifted vertically
in the order of descending energy from top to bottom of the figure.
}
\end{figure}

In this paper, we assume that the number of electrons in the 2D topological insulator will not be strongly
affected by the introduction of lateral superlattice potential.  This approximation is consistent with the
assumption of diagonal addition of the superlattice potential. If the fermi energy is near the band gap of the bulk
2D topological insulator which is taken as the energy zero point, then in the presence of the lateral superlattice
potential the fermi energy should be close to the averaged potential.  Thus in this paper we will focus on the
bands near the averaged potential.  For convenience we will label six bands near the averaged potential as $E1$,
$E2$, $E3$, $H1$, $H2$ and $H3$, respectively (see Figure 1).

In Figure 2(a), we show the energy dispersion along the $q_x$ direction in the first Brillouin zone
($[-\pi/L, \pi/L]$) for different values of $V_0$ and $k_y=0$. 
\begin{figure}[ht]
\includegraphics[width=6cm,height=9cm,angle=-90]{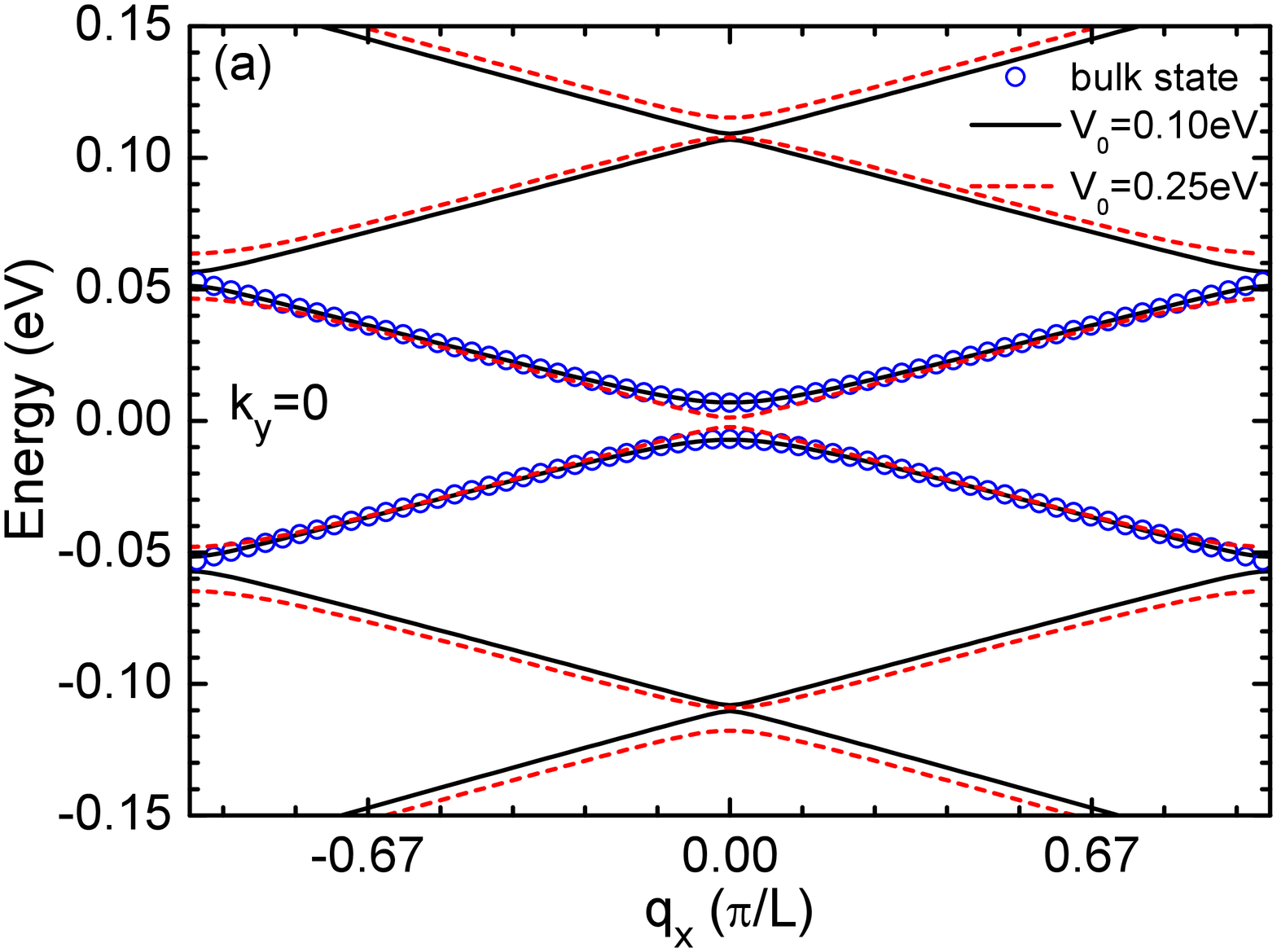}
\includegraphics[width=6cm,height=9cm,angle=-90]{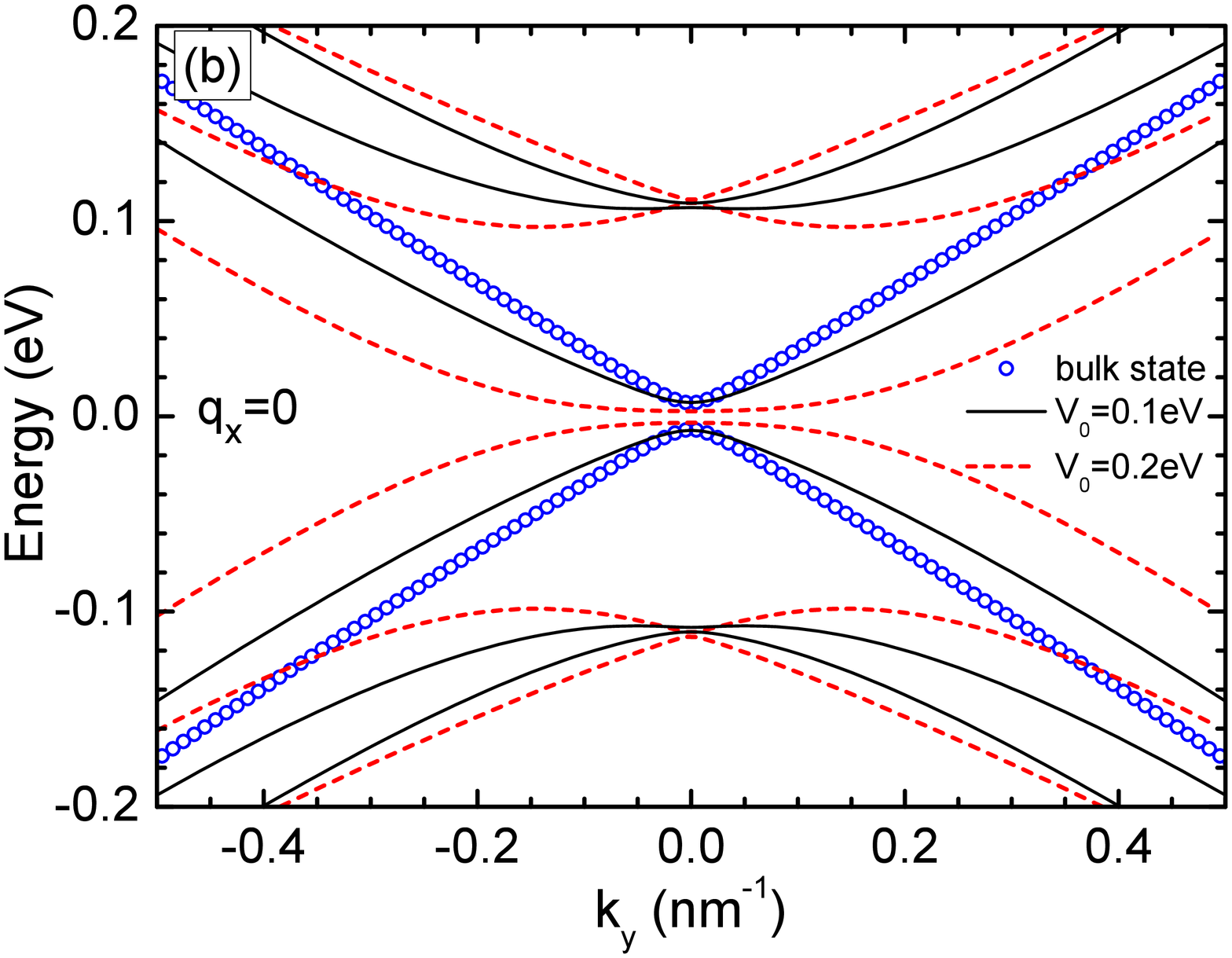}
\caption{(Color online)
(a) The energy dispersion along the $q_x$ direction in the first Brillouin zone
of the lateral superlattice with $k_y=0$.  The averaged potential energy $V_0R/L$
is subtracted from each energy level.
The black solid line is for $V_0=0.1$ eV and the red dashed line is for $V_0=0.25$ eV.
The open circles show the energy dispersion of the bulk state with no shift.
(b) The energy levels versus $k_y$ with the averaged potential energy
subtracted for $q_x=0$.  The black solid line is for $V_0=0.1$ eV
and the red dashed line is for $V_0=0.2$ eV.
The open circles show the energy dispersion of the bulk state.
}
\end{figure}
In Figure 2(b), the energy
dispersion along the $k_y$ direction is shown with $q_x=0$. In each energy dispersion, the averaged
potential energy $V_0 R/L$ is subtracted. The bulk bands are also shown in Figure 2 as open
symbols.  When $V_0$ is small, the shape of bands $E1$ and $H1$ are close to the bulk bands as
expected. We wish to point out that when the sign of $M$ is reversed, the energy dispersion is
qualitatively unchanged.

The energy dispersion changes in a rather complicated way when the barrier potential $V_0$ changes.
In Figure 3, six energy levels are shown as a function of $V_0$, with the average potential
subtracted, for various values of $q_x$ and a fixed $k_y=0$.  One can clearly see that all bands
display an oscillatory behavior.  There are three critical values of the potential $V_0$ at which
the $E1$ and $H1$ bands (E2 and E3, H2 and H3) touch each other when $q_x=k_y=0$. The oscillation
amplitude will be smaller and no touching point exists when $q_x\ne 0$ or $k_y\ne 0$.
\begin{figure}[ht]
\includegraphics[width=6cm,height=9cm,angle=-90]{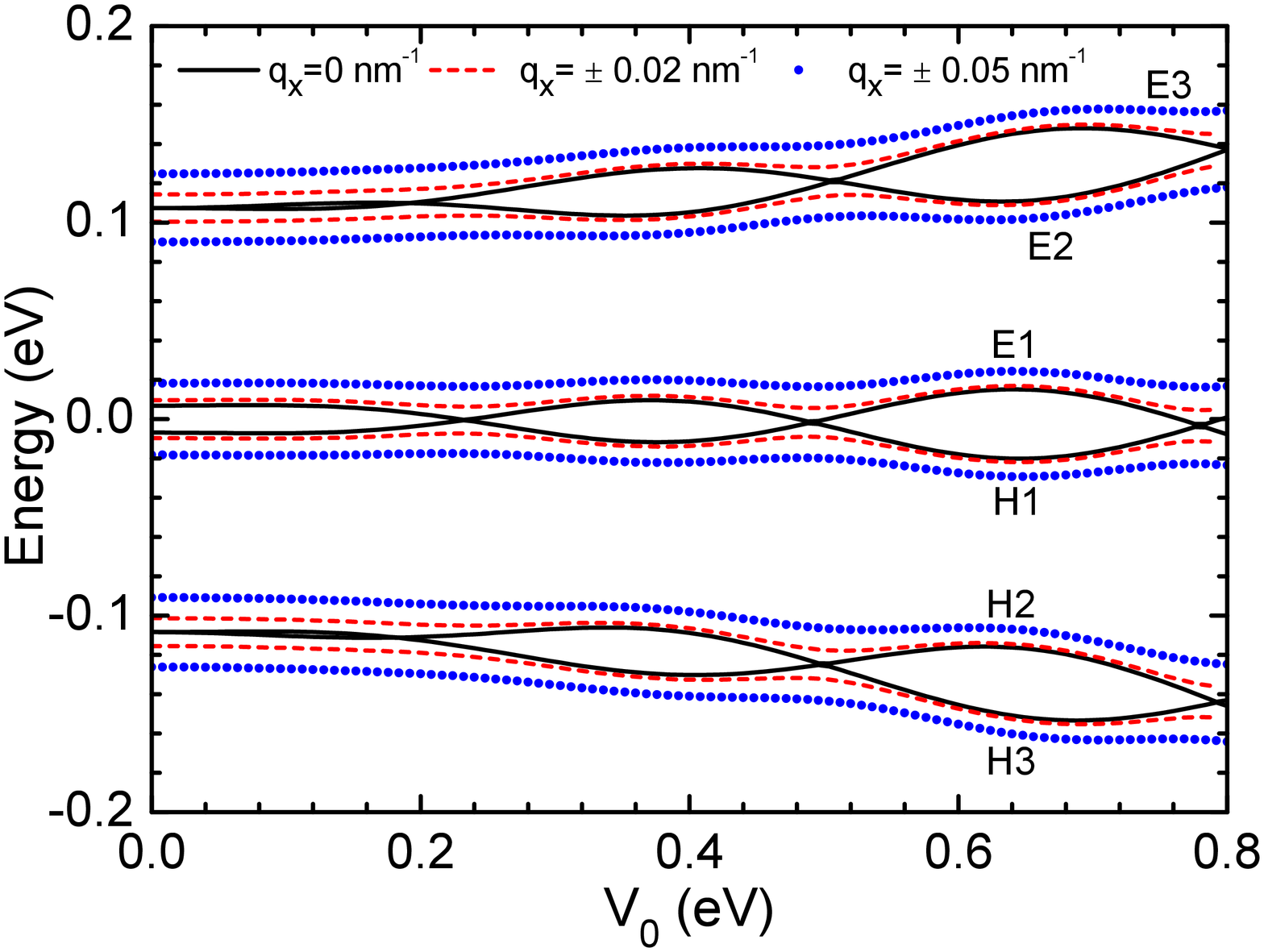}
\caption{(Color online)
Energy levels versus $V_0$ the lateral superlattice barrier potential amplitude
for various values of $q_x$ and a fixed $k_y=0$.
The averaged potential energy is subtracted from each energy level.
Curves with opposite sign of $q_x$ are identical.
}
\end{figure}

Next, we study the spatial dependence of wave functions.  The system considered is uniform along
the $y$ direction, therefore we only need to investigate the probability distribution along the $x$
direction.  The total distribution, for a specific band, can be written as $|\Psi(x)|^2=|u^{\rm
u}(x)|^2+|u^{\rm l}(x)|^2$ contributed from the upper and lower components of the corresponding
wave function. Figure 4 shows the probability distribution for various value of $q_x$ and $k_y$
with a fixed $V_0=0.2$ eV.  The well/barrier regions are indicated by two vertical lines in Figure
4.  Figure 4(a) is for the $E1$ band and Figure 4(b) for the $H1$ band. It can be seen that, for
the $E1$ band, the amplitude of wave function in the middle well region ($R/2<x<L-R/2$) is larger
than that in the barrier region ($0<x<R/2$ and $R/2<x<L$).  But for the $H1$ band, the amplitude of
wave function is larger in the barrier region.
\begin{figure}[ht]
\includegraphics[width=6cm,height=9cm,angle=-90]{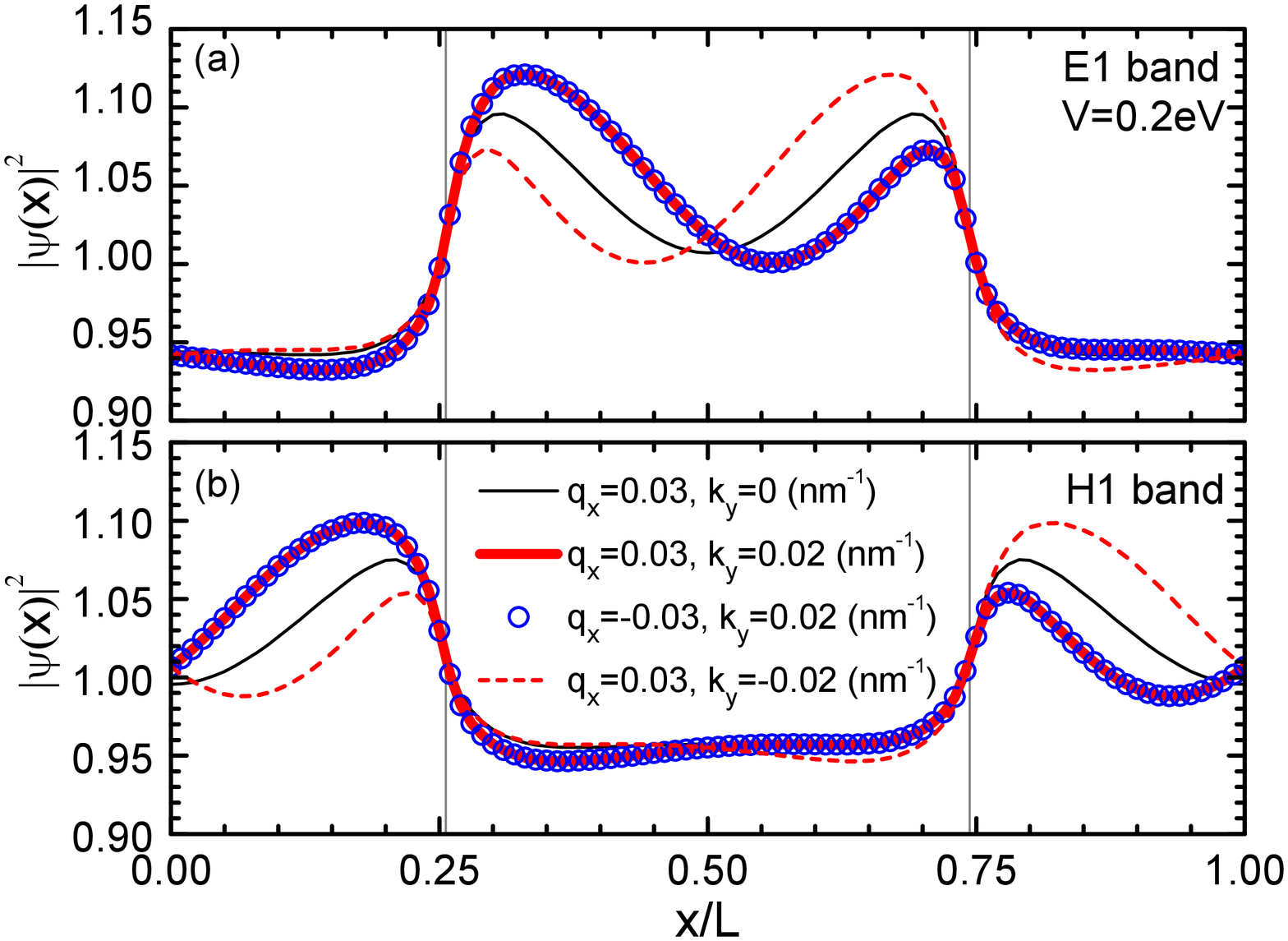}
\caption{(Color online)
The spatial dependence of the total wave function in one lateral superlattice unit
cell.  (a) The total wave function of $\rm E1$ band and (b) that of $\rm H1$ band.
}
\end{figure}

It is interesting to note that the energy of $E1$ and $H1$ bands is about $0.1$ eV, both below the
barrier potential $V_0=0.2$ eV.  On the other hand, the distribution shows that the $E1$ band
behaves like a state in the well, while the $H1$ band behaves like a state in the barrier, if we
view them in a traditional single band picture.  In the single band case, when the band energy is
below the barrier of a superlattice potential, the wave function should be more confined in the
well region.  The result shown in Figure 4 indicates that a state whose energy is below the barrier
could become ``localized'' in the well region or in the barrier region.  This resembles the case of
above-barrier states in a semiconductor superlattice \cite{PhysRevB.38.9667, PhysRevLett.68.3220,
PhysRevB.47.3806, PhysRevB.64.195311}. This effect is not unexpected, as in the present study, we
deal with a two-band model with a strong inter-band coupling.

As shown in Figure 4, electronic states with $k_y=0$ have a symmetrical distribution about the
center of the unit cell $x=L/2$.  This symmetry will be destroyed if $k_y\ne 0$. However, the
distribution of a state with $k_y=k_y^0\ne 0$ and the distribution of a state with $k_y=-k_y^0$ are
connected by a reflection transformation about the center of the unit cell.  By making the
reflection transformation, one obtains the distribution for $k_y=-k_y^0$ from the distribution for
$k_y=k_y^0$ and vice versa. This symmetry about the distribution is due to the symmetry of the
model Hamiltonian that $h^*(k_x,k_y)=h(k_x,-k_y)$.

Next, we study how the probability distribution (the wave function) changes when the lateral
superlattice potential $V_0$ changes.  This task becomes easier if we introduce the notion of
probability distribution integrated over the well region in an unit cell $P_{\rm w}=P_{\rm w}^{\rm
u}+P_{\rm w}^{\rm l}$ with $P^{\rm u}_{\rm w}=\int_{R/2}^{L-R/2} |u^{\rm u}(x)|^2 dx$ and $P^{\rm
l}_{\rm w}=\int_{R/2}^{L-R/2} |u^{\rm l}(x)|^2 dx$. Similarly, we also introduce $P_{\rm b}$ the
probability distribution integrated over the barrier region in an unit cell.  In Figure 5, we show
the total $P_{\rm w}$ for the $E1$ and $H1$ bands for various values of $q_x$, $k_y$ and $M$.  It
is clear that, as the barrier potential $V_0$ increases, the probability distribution in the well
displays an oscillatory behavior around $0.5$ the homogenous value.  When $q_x=0$, the oscillation
of $P_{\rm w}$ shows abrupt changes at several values of $V_0$ (see Figure 5(a) and also Figure
5(d)).  When $q_x\ne 0$, these abrupt changes becomes smoother.
\begin{figure}[ht]
\includegraphics[width=6cm,height=9cm,angle=-90]{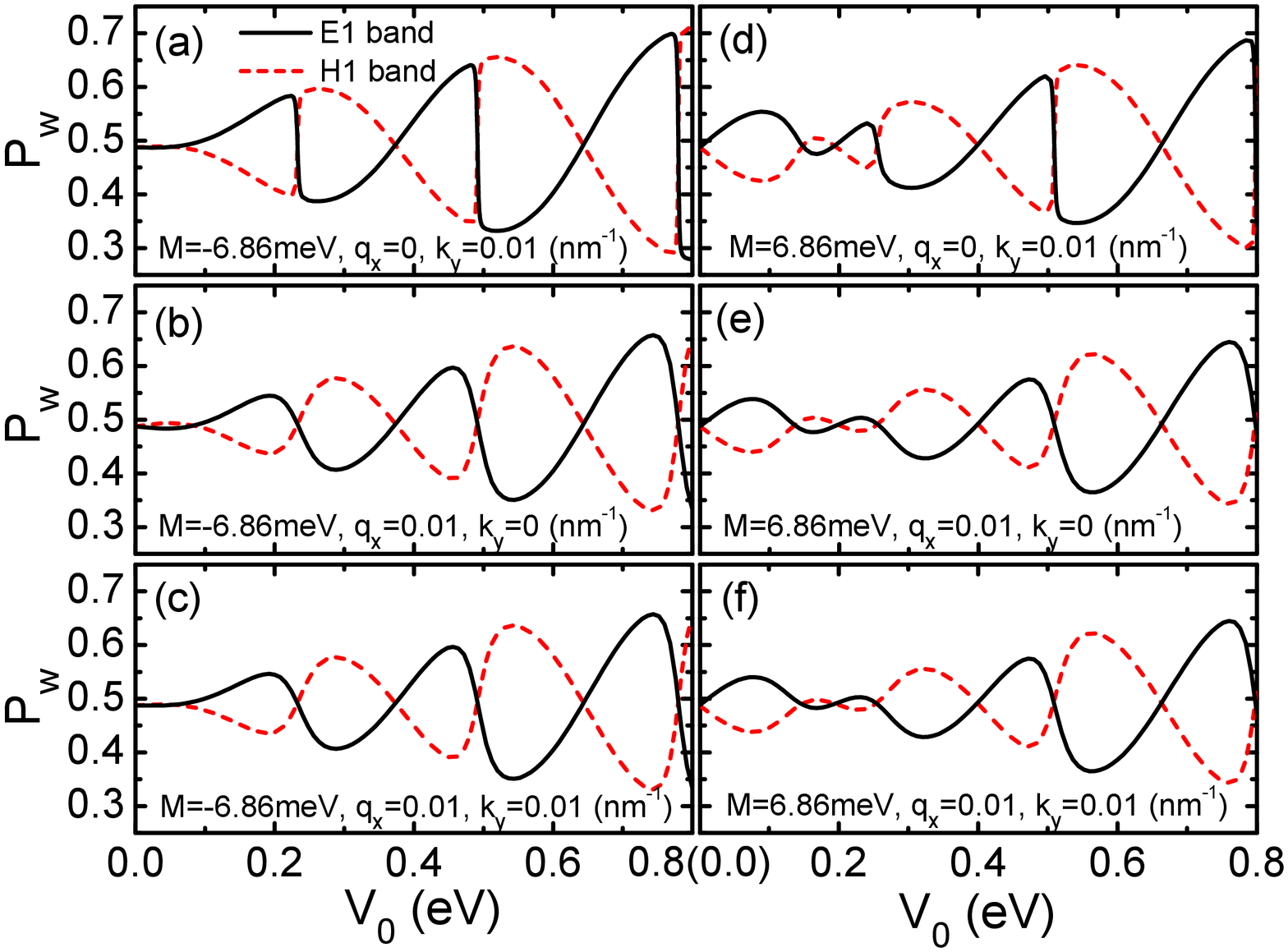}
\caption{(Color online)
The wave function integrated in the well region $P_{\rm w}$ versus $V_0$ for various
values of $q_x$ and $k_y$.  The solid curves are for the $\rm E1$ band and dashed
curves are for the $\rm H1$ band.  In (a), (b), and (c) $M=-6.86$ meV and in (d), (e),
and (f) $M=6.86$ meV.
In (a) and (d), $q_x=0$, $k_y=0.01$ nm$^{-1}$.   In (b) and (e), $q_x=0.01$ nm$^{-1}$, $k_y=0$.
In (c) and (f), $q_x=0.01$ nm$^{-1}$, $k_y=0.01$ nm$^{-1}$.
}
\end{figure}

The integrated probability distribution for the $E2$, $E3$, $H2$, and $H3$ bands are shown in
Figure 6 for $q_x=0.01$ nm$^{-1}$, $k_y=0$ and two values of $M$.  
\begin{figure}[ht]
\includegraphics[width=6cm,height=9cm,angle=-90]{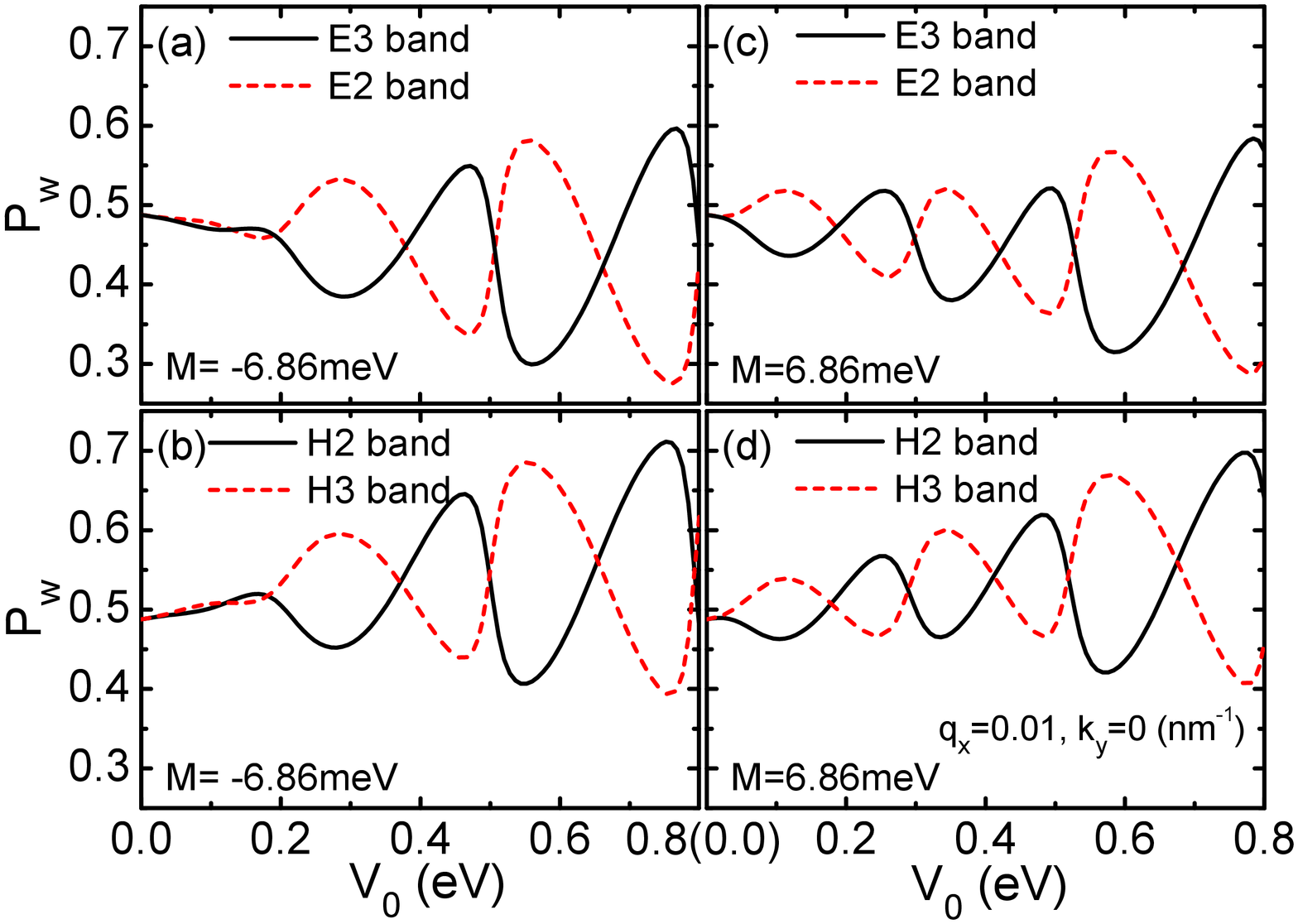}
\caption{(Color online)
The wave function integrated in the well region $P_{\rm w}$ versus $V_0$ for $q_x=0.01$ nm$^{-1}$
and $k_y=0$.  Panels (a) and (c) are for $\rm E2$ and $\rm E3$ bands, panels (b) and (d) are
for $\rm H2$ and $\rm H3$ bands.  In (a) and (b) $M=-6.86$ meV and in (c) and (d) $M=6.86$ meV.
}
\end{figure}
The oscillatory behavior is
similar to that of $E1$ and $H1$ bands except that the oscillation is slightly slantwise to the
homogeneous distribution.  States with even higher/lower energies are also investigated and similar
results are found.  This result is consistent with that in Reference \cite{PhysRevLett.106.057205},
where the author found that the spin polarization oscillates across the boundary of a step-function
electrical potential on the surface of a three dimensional topological insulator and is not
confined by the potential due to the Klein paradox.  By examining Figure 5 and 6, one observes that
the amplitude of the oscillation is not strong and this indicates that the effectiveness of gating
on top of a 2D topological insulator is limited.  As we have mentioned, this is due to the coupled
two-band nature intrinsic to the system.  Moreover, it should be pointed out that the oscillatory
pattern can be modified dramatically when a different set of topological insulator parameters
\cite{PhysRevLett.101.246807} is used.

It is well-known that, when the effective model for the 2D topological insulator is applied to a
stripe geometry system, the existence of the nontrivial edge states requires that the bulk bands
must be inverted, i.e., $M<0$ \cite{PhysRevLett.101.246807}.  When all parameters except $M$ are
unchanged, the model with negative $M$ and positive $M$ have completely different behavior when
applied to the stripe system.  In Figure 5 and 6, we compare $P_{\rm w}$ calculated with a negative
$M$ and a positive one. The left panels are for the negative $M$ case and the right panels are for
the positive $M$ case. In the case of positive $M$, one observes that the number of oscillations
increases. However, it is clear that there is no significant difference between two cases.
\begin{figure}[ht]
\includegraphics[width=6cm,height=9cm,angle=-90]{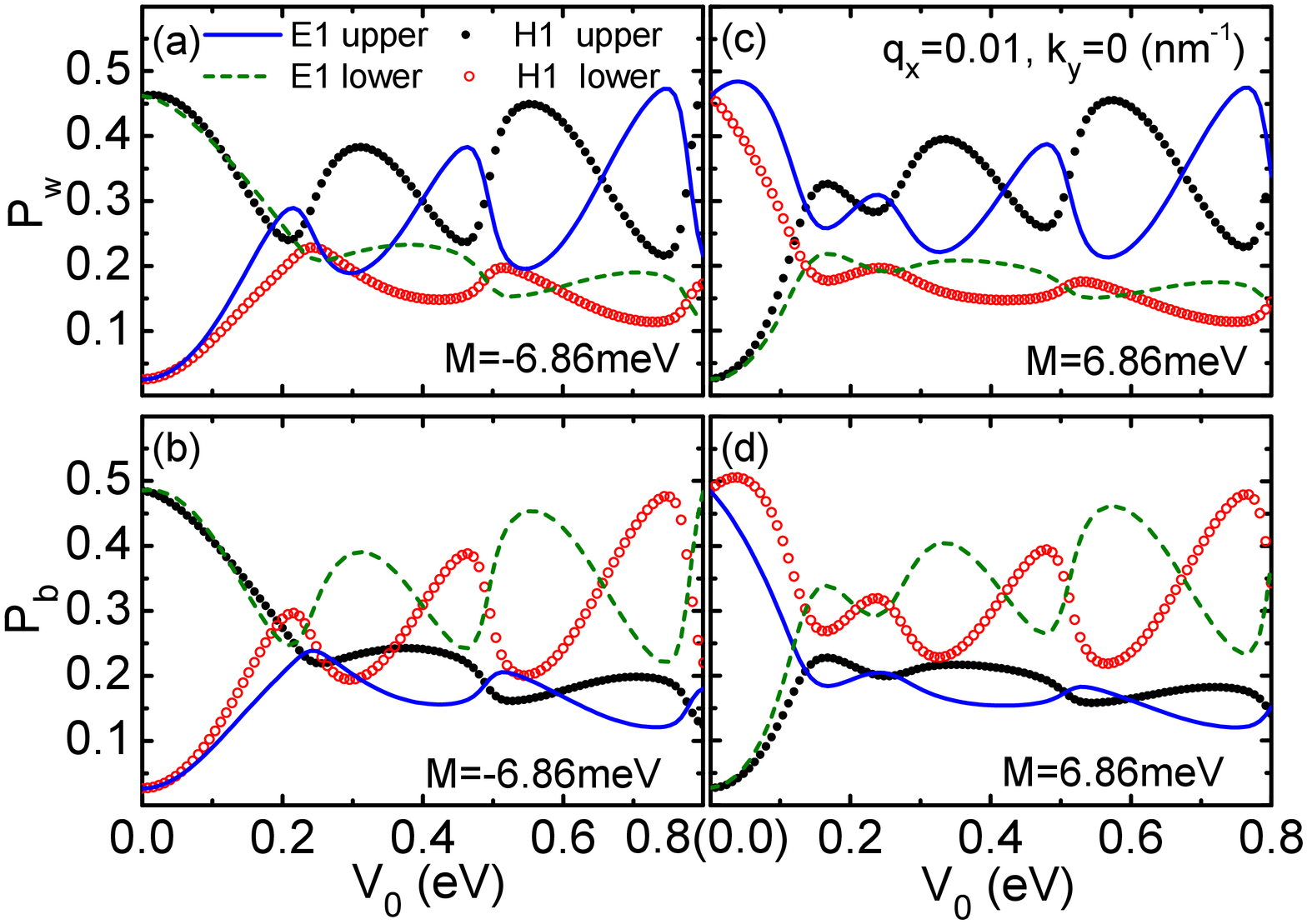}
\caption{(Color online)
Panels (a) and (c) show the wave function integrated in the well region $P_{\rm w}$ versus $V_0$,
for the upper and lower components of the wave function.
Panels (b) and (d) show the wave function integrated in the barrier region
versus $V_0$ for both upper and lower components.
Only the $\rm E1$ and $\rm H1$ bands represented.
$q_x=0.01$ nm$^{-1}$ and $k_y=0$.
In (a) and (b) $M=-6.86$ meV and in (c) and (d) $M=6.86$ meV.
}
\end{figure}

As the lateral superlattice studied in this paper is a coupled two-band system, we have evaluated
the probability distribution for each component of a state. In Figure 7, the integrated probability
distribution (wave function) is shown for the upper and lower component of wave function
separately.  The integral is performed either inside the potential well region (Figures 7(a) and
7(c)) or in the barrier region (Figures 7(b) and 7(d)). Only the results for the $E1$ and $H1$
bands are presented.  In the left panels $M<0$, and in the right panels $M>0$.  $q_x=0.01$
nm$^{-1}$ and $k_y=0$.  One observes that the two components both display oscillatory behavior.

In the case of $M<0$ (Figures 7(a) and 7(b)), when $V_0$ is small, $P_{\rm w}$ and $P_{\rm b}$ of
the $H1$ band mainly consists of the upper component, and for the $E1$ band they are dominated by
the lower component.  As $V_0$ increases, the difference between two components becomes small.  In
contrast, when $M>0$ (Figures 7(c) and 7(d)) and $V_0$ is small, the $E1$ band is mainly
contributed from the upper component and the $H1$ band is dominated by the lower component. More
calculations show that when $q_x$ and $k_y$ becomes large, the contribution to $P_{\rm w}$ and
$P_{\rm b}$ from upper and lower components will become nearly the same, even in the region of
small $V_0$.

%
%

\section{Summary}

In summary, we have investigated theoretically the properties of electronic states of a lateral
superlattice made of a 2D topological insulator.  The dispersion of mini-band and the electron wave
function in the potential well region are found to display an oscillatory behavior as the potential
strength of the lateral superlattice increases. The probability of finding an electron in the
potential well region can be larger or smaller than the average as the potential strength varies.
This indicates that the electrical confining effect in two-dimensional topological insulators is
not high and one should be careful with the idea of creating boundaries using gate voltage in
device design.  These features are different from that in a single-band Kronig-Penney model, and
can be attributed to the coupled-multiple-band nature of the 2D topological insulator model.

%
%

\begin{acknowledgments}

This work was partly supported by NSF and MOST of China.

\end{acknowledgments}

%
%

%
%
%

\end{document}